\documentclass[conference]{IEEEtran}

\usepackage{physics}
\usepackage{graphicx}
\usepackage{grffile}
\usepackage{amsmath}
\usepackage{mathtools}
\usepackage{autobreak}
\usepackage{amssymb}
\usepackage{amsfonts}
\usepackage{multirow}
\usepackage{hyperref}

\begin{document}

\title{Symmetric Ternary Logic and Its Systematic Logic Composition Methodology}

\author{
  \IEEEauthorblockN{
    Ichiro Kawashima, Member, IEEE
  }
  \IEEEauthorblockA{
    ichiro.kawashima@ieee.org
  }
}

\maketitle

\begin{abstract}
Ternary logic is expected to increase the area efficiency of VLSI due to its expressiveness compared to the traditional binary logic.
This paper proposes a new symmetric ternary logic and a systematic logic composition methodology that enables us to design any ternary logic circuits.
The methodology is demonstrated by implementing the ternary inverters, ternary NAND, ternary NOR, and ternary half-adder operators with the proposed symmetric ternary operators.
\end{abstract}


\section{Introduction}

The development of digital circuits based on binary logic produced significant progress in the information technology.
The systematic methodology of designing digital circuits has been established and developed.
The systematic methodology makes it possible to express any data and calculation on digital circuits with two values (i.e. $0$ and $1$).
However, it is known that the microfabrication of semiconductors following Moore's law is reaching its end.
The end of Moore's law results in the stagnation of the progress of the information technology which is based on digital hardware.

Ternary logic is expected to contribute to the progress of the information technology after the end of Moore's law by using three values (i.e. $0$, $1$, and $2$) instead of two values.
Three-valued logic is potentially more expressive than two-valued logic; for example, binary logic has $4 ~ (= 2 ^ 2)$ monadic operators and $16 ~ (= 2 ^ 4)$ dyadic, meanwhile, ternary logic has $27 ~ (= 3 ^ 3)$ possible monadic operators and $19,683 ~ (= 3 ^ 9)$ possible dyadic operators.
The expressive operators can reduce a large number of connections in VLSI.

There is a considerable number of studies on ternary logic.
Most studies on implementations of ternary logic are based on Kleene's logic.
In his logic, the binary logic which is composed of the true, false, logical conjunction, and logical disjunction is extended, and the third value is recognized as an intermediate (unknown) value between the true and false~\cite{fitting1991kleene}.
The symmetrical relationship between the true and false is inherited from the binary logic but the third value is not handled equally to the former two values in this logic.


Ternary logics based on Kleene's logic are implemented into physical devices, such as complementary metal-oxide semiconductors (CMOS) circuits~\cite{JAHANGIR201982,saha2020systematic} and carbon nano-tube field effect transistors (CNFET)~\cite{sharma2020cnfet,lin2009cntfet,jaber2019high}.
In those studies, there are many ternary operators introduced to calculate ternary values, such as a standard ternary inverter (STI), negative ternary inverter (NTI), positive ternary inverter (PTI), ternary NAND (TNAND), ternary NOR (TNOR), ternary decoder (TDecoder), and ternary half-adder (THA)~\cite{jaber2019high}.

In this paper, new ternary logic that focuses on the symmetry of the three values is proposed.
This work revealed that only one monadic operator and three dyadic operators are required to express any ternary calculation.
Moreover, a systematic logic composition methodology that enables us to design any desired ternary logic circuits is introduced.

\section{Methods}

This paper proposed four operators for the ternary calculation, which is composed of one monadic operator and three dyadic operators.
In this paper, the binary algebra is extended for ternary values with preserving its symmetric structure.
In this section, the conventional binary logic is reviewed and a new ternary logic is introduced.

\subsection{Traditional Binary Logic}

\begin{table}[h]
    \centering
    \begin{tabular}{cc}
        \begin{minipage}{0.3\linewidth}
            \centering
            \caption{Truth table of the NOT operator} \label{tab:binary_not}
            \begin{tabular}{c|c}
                    $X$ & $\overline{X}$ \\
            \hline 
                    $0$ & $1$            \\
                    $1$ & $0$            \\
            \end{tabular}
        \end{minipage} &
        \begin{minipage}{0.6\linewidth}
            \centering
            \caption{Truth table of the AND and OR operators} \label{tab:binary_and_or}
            \begin{tabular}{cc|cc}
                    $X$ & $Y$ & $X \times Y$ & $X + Y$ \\
            \hline 
                    $0$ & $0$ & $0$          & $0$     \\
                    $0$ & $1$ & $0$          & $1$     \\
                    $1$ & $0$ & $0$          & $1$     \\
                    $1$ & $1$ & $1$          & $1$     \\
            \end{tabular}
        \end{minipage}
    \end{tabular}
\end{table}

The truth table of the monadic operator and dyadic operators are shown in Table~\ref{tab:binary_not} and Table~\ref{tab:binary_and_or}.
The NOT ($\overline{X}$) operator changes $0$ to $1$ and $1$ to $0$.
The operator can be interpreted as a symmetrical transformation of $0$ and $1$, where $0$ and $1$ are continuous with each other.
The AND ($X \times Y$) operator precedes $0$ and outputs $0$ unless both of the two inputs are $1$.
Likewise, the OR ($X + Y$) operator precedes $1$ and outputs $1$ unless both two inputs are $0$.

\begin{table}[h]
    \centering
    \caption{Laws stand up on the binary logic} \label{tab:binary_laws}
    \begin{tabular}{cl}
        \multirow{1}{*}{Involution}      & $\overline{\overline{x}} = x \text{.}$                \\
        \hline 
        \multirow{2}{*}{Boundedness}     & $x \times 0 = 0$                                      \\
                                         & $x + 1 = 1$                                           \\
        \hline 
        \multirow{2}{*}{Identity}        & $x \times 1 = x$                                      \\
                                         & $x + 0 = x$                                           \\
        \hline 
        \multirow{2}{*}{Complementation} & $x \times \overline{x} = 0$                           \\
                                         & $x + \overline{x} = 1$                                \\
        \hline 
        \multirow{2}{*}{Idempotency}     & $x \times x = x$                                      \\
                                         & $x + x = x$                                           \\
        \hline 
        \multirow{2}{*}{Commutativity}   & $x \times y = y \times x$                             \\
                                         & $x + y = y + x$                                       \\
        \hline 
        \multirow{2}{*}{Associativity}   & $(x \times y) \times z = x \times (y \times z)$       \\
                                         & $(x + y) + z = x + (y + z)$                           \\
        \hline 
        \multirow{2}{*}{Distributivity}  & $x \times (y + z) = (x \times y) + (x \times z)$      \\
                                         & $x + (y \times z) = (x + y) \times (x + z)$           \\
        \hline 
        \multirow{2}{*}{De Morgan}       & $\overline{x \times y} = \overline{x} + \overline{y}$ \\
                                         & $\overline{x + y} = \overline{x} \times \overline{y}$ \\
    \end{tabular}
\end{table}

Table~\ref{tab:binary_laws} illustrates the laws the binary logic fulfills.
The involution law tells any binary values (i.e. $0$ and $1$) can be restored by applying the NOT operator twice because the values are transformed into the other value symmetrically at the first application of the NOT, and transformed back into the original value at the second application.
The boundedness law explains that $0$ is the bound of the OR operator and $1$ is the bound of the AND operator; any value is trapped into the bound.
The identity law describes that the AND operator can be canceled by applying $1$, and the OR operator can be canceled by applying $0$, and so can applying two same values to the dyadic operators (idempotency).
The complementation law shows any binary value and its inverted value can make a constant value by applying them to dyadic operators together.
Moreover, swapping two inputs or changing the application order of the dyadic operators does not change the calculation result (commutativity and associativity).
The calculation order of the AND and OR operators can be reversed as described in the distributivity law.
De Morgan's duality tells that a logical expression and its transformed expression that swap $0$ and $1$ and swap the AND and OR operators can be equated.

The regular formula of monadic operators of binary logic is defined as follows:
\begin{align}
    \begin{autobreak}
        f (a_0, a_1)
            =  \overline{x} \times a_0
            +            x  \times a_1
        \end{autobreak} \label{equ:binary_lut1}
\end{align}
$4 ~ (= 2 ^ 2)$ possible monadic operators can be defined by putting $0$ or $1$ to $a_0$ and $a_1$ in the formula.
For instance, the formula becomes the identity and NOT if $(a_0, a_1)$ is $(0, 1)$ and $(1, 0)$, respectively.
the formula becomes the constant $0$ if $(a_0, a_1)$ is $(0, 0)$ due to the boundedness of $0$ and OR, and the formula becomes constant $1$ for $(1, 1)$ owing to the identity of $1$ and AND, and the complementation of $1$ and the OR operator. 

The regular formula of dyadic operators of binary logic refers as follows:
\begin{align}
    \begin{autobreak}
    f (a_0, \cdots, a_3)
        =       \overline{x} \times \overline{y} \times a_0
        +       \overline{x} \times           y  \times a_1
        +                 x  \times \overline{y} \times a_2
        +                 x  \times           y  \times a_3
    \end{autobreak} \label{equ:binary_lut2}
\end{align}
The formula describe $16 ~ (= 2 ^ {2 \times 2})$ possible dyadic operators by setting $0$ or $1$ to $a_0$ to $a_3$.
As described in the monadic operators, combinations of $a_0$ to $a_3$ make the formula become dyadic binary operators.
For instance, the AND, OR, XOR, NAND, and NOR operators can be obtained when $(a_0, a_1, a_2, a_3)$ is set as $(0, 0, 0, 1)$, $(0, 1, 1, 1)$, $(0, 1, 1, 0)$, $(1, 1, 1, 0)$, $(1, 0, 0, 0)$, respectively.

\begin{table}[h]
    \centering
    \begin{tabular}{cc}
        \begin{minipage}{0.4\linewidth}
            \centering
            \caption{Truth table of binary monadic operators} \label{tab:binary_lut1}
            \begin{tabular}{c|c}
                    $X$ & $A$   \\
            \hline 
                    $0$ & $a_0$ \\
                    $1$ & $a_1$ \\
            \end{tabular}
        \end{minipage} &
        \begin{minipage}{0.4\linewidth}
            \centering
            \caption{Truth table of binary dyadic operators} \label{tab:binary_lut2}
            \begin{tabular}{cc|c}
                    $X$ & $Y$ & $A$   \\
            \hline 
                    $0$ & $0$ & $a_0$ \\
                    $0$ & $1$ & $a_1$ \\
                    $1$ & $0$ & $a_2$ \\
                    $1$ & $1$ & $a_3$ \\
            \end{tabular}
       \end{minipage}
    \end{tabular}
\end{table}

Those regular formulae make it possible to construct operators from truth tables as shown in Table~\ref{tab:binary_lut1}, and Table~\ref{tab:binary_lut2}.
The binary operators can be designed by putting the desired output values for each input value on the truth tables and replacing $a_0, \cdots, a_3$ with the output values.

\subsection{Proposed Ternary Logic}

\begin{table}[h]
    \centering
    \begin{tabular}{cc}
        \begin{minipage}{0.3\linewidth}
            \centering
            \caption{Truth table of the ROTATE operator} \label{tab:ternary_rotate}
            \begin{tabular}{c|c}
                    $X$ & $\overline{X}$ \\
            \hline 
                    $0$ & $2$            \\
                    $1$ & $0$            \\
                    $2$ & $1$            \\
            \end{tabular}
        \end{minipage} &
        \begin{minipage}{0.6\linewidth}
            \centering
            \caption{Truth table of the ALPHA, BETA and GAMMA operator} \label{tab:ternary_alpha_beta_gamma}
            \begin{tabular}{cc|ccc}
                    $X$ & $Y$ & $X \times Y$ & $X + Y$ & $X \diamond Y$ \\
            \hline 
                    $0$ & $0$ & $0$          & $0$     & $0$            \\
                    $0$ & $1$ & $0$          & $1$     & $0$            \\
                    $0$ & $2$ & $0$          & $2$     & $2$            \\
                    $1$ & $0$ & $0$          & $1$     & $0$            \\
                    $1$ & $1$ & $1$          & $1$     & $1$            \\
                    $1$ & $2$ & $1$          & $1$     & $2$            \\
                    $2$ & $0$ & $0$          & $2$     & $2$            \\
                    $2$ & $1$ & $1$          & $1$     & $2$            \\
                    $2$ & $2$ & $2$          & $2$     & $2$            \\
            \end{tabular}
       \end{minipage}
    \end{tabular}
\end{table}

The truth table of the monadic operator and dyadic operators are shown in Table~\ref{tab:ternary_rotate} and Table~\ref{tab:ternary_alpha_beta_gamma}.
As shown in the tables, the ROTATE ($\overline{X}$) operator changes $0$ to $2$, $1$ to $0$, and $2$ to $1$ and transformed ternary values symmetrically as the NOT operator of the binary logic does.
As the AND and OR operators of the binary logic precede $0$ and $1$, the ALPHA ($X \times Y$), BETA ($X + Y$), and GAMMA ($X \diamond Y$) operators in my proposal precede $0$, $1$, and $2$, respectively.
If the precedent value does not appear in the operation, the operators precede the secondary precedent value.
For instance, the ALPHA operator precedes $1$ if the input is $1$ or $2$; the BETA operator precedes $2$ if the input is $2$ or $0$; the GAMMA operator precedes $0$ if the input of the operator is $0$ and $1$.
Note that the ALPHA and BETA operators work as the AND and OR operators in the traditional binary logic if the input of those operator does not includes $2$.
The calculation priority of the ternary operations is the ROTATE, ALPHA, BETA, and GAMMA operators if parentheses are omitted in this paper.

\begin{table}[h]
    \centering
    \caption{Laws stand up on the ternary logic} \label{tab:ternary_laws}
    \begin{tabular}{cl}
        \multirow{1}{*}{Involution}      & $\overline{\overline{\overline{x}}} = x$                         \\
        \hline 
        \multirow{3}{*}{Boundedness}     & $x \times 0 = 0$                                                 \\
                                         & $x + 1 = 1$                                                      \\
                                         & $x \diamond 2 = 2$                                               \\
        \hline 
        \multirow{3}{*}{Identity}        & $x \times 2 = x$                                                 \\
                                         & $x + 0 = x$                                                      \\
                                         & $x \diamond 1 = x$                                               \\
        \hline 
        \multirow{3}{*}{Complementation} & $x \times \overline{x} \times \overline{\overline{x}} = 0$       \\
                                         & $x + \overline{x} + \overline{\overline{x}} = 1$                 \\
                                         & $x \diamond \overline{x} \diamond \overline{\overline{x}} = 2$   \\
        \hline 
        \multirow{3}{*}{Idempotency}     & $x \times x = x$                                                 \\
                                         & $x + x = x$                                                      \\
                                         & $x \diamond x = x$                                               \\
        \hline 
        \multirow{3}{*}{Commutativity}   & $x \times y = y \times x$                                        \\
                                         & $x + y = y + x$                                                  \\
                                         & $x \diamond y = y \diamond x$                                    \\
        \hline 
        \multirow{3}{*}{Associativity}   & $(x \times y) \times z = x \times (y \times z)$                  \\
                                         & $(x + y) + z = x + (y + z)$                                      \\
                                         & $(x \diamond y) \diamond z = x \diamond (y \diamond z)$          \\
        \hline 
        \multirow{3}{*}{Distributivity}  & $x \times (y + z) = (x \times y) + (x \times z)$                 \\
                                         & $x + (y \diamond z) = (x + y) \diamond (x + z)$                  \\
                                         & $x \diamond (y \times z) = (x \diamond y) \times (x \diamond z)$ \\
        \hline 
        \multirow{3}{*}{De Morgan}       & $\overline{x \times y} = \overline{x} \diamond \overline{y}$     \\
                                         & $\overline{x + y} = \overline{x} \times \overline{y}$            \\
                                         & $\overline{x \diamond y} = \overline{x} + \overline{y}$          \\
    \end{tabular}
\end{table}

Table~\ref{tab:ternary_laws} shows the laws the ternary logic satisfies.
The proposed ternary operations are designed to fulfill those laws as binary operators.
The ROTATE operator can transform the input symmetrically, and the input value can be restored by applying ROTATE three times.
The bounds of the ALPHA and BETA operators are $0$ and $1$, just like the AND and OR operators in the traditional binary logic, and $2$ is the bound of the GAMMA operator.
The ternary logic in this work also has the identity law.
The complementation law takes three terms and each term is transformed by the ROTATE operator; a constant ternary value is obtained by applying those terms to the same dyadic operator.
The dyadic ternary operators do not change the input value if the two input values are the same (idempotency).
Moreover, my proposal also fulfills the commutativity and associativity laws.
The distributivity law is satisfied in the ALPHA and BETA operators if the BETA operator is calculated first; the BETA and GAMMA operators if the GAMMA operator is calculated first; the GAMMA and ALPHA operators if the ALPHA operators are calculated first.
However, the distributivity law does not stand on the operation when the calculation order is reversed (e.x. $x + (y \times z) \neq (x + y) \times (x + z)$).
Finally, my symmetrical design of ternary operators enables us to transform ternary logical expressions by De Morgan's law as shown in the table.

The regular formula of monadic operators of ternary logic is defined as follows:
\begin{align}
    \begin{autobreak}
        f (a_0, a_1, a_2)
            =                            x   \times 1 + a_0
            \diamond           \overline{x}  \times 1 + a_1
            \diamond \overline{\overline{x}} \times 1 + a_2
    \end{autobreak} \label{equ:ternary_lut1}
\end{align}
Just like (\ref{equ:binary_lut1}), putting $0$, $1$ or $2$ to $a_0, \cdots, a_8$ lets the formula behave as $27 ~ (= 3 ^ 3)$ possible monadic operators.
In the formula, a constant ternary value $1$ appears unlike the regular formula in binary logic.
The part $\times 1$ works as a filter that passes $0$ and $1$ and merges $2$ into $1$; therefore, $x \times 1$ only becomes $0$ or $1$ for any ternary value of $x$.
The part $x \times 1 + a_0$ becomes $a_0$ if $x \times 1 = 0$ because of the identity law of the operator $+$.
The part $x \times 1 + a_0$ becomes $1$ for any value of $a_0$ if $x \times 1 = 1$ because $1$ is the bound of the operator $+$.
$1 \diamond$ works as an identity function; hence, the term $x \times 1 + a_0$ burnishes if $x \times 1 = 1$.
In short, two of three terms in (\ref{equ:ternary_lut1}) banish for any input of $x$ due to the boundedness of $+$ and the identity of $\diamond$, then the rest of the term becomes $a_0$, $a_1$ or $a_2$ owing to the identity of $+$.
One of $a_0$, $a_1$ and $a_2$ is selected by the input $x$ in the formula, just like either of $a_0$ or $a_1$ is selected by the input $x$ in the formula (\ref{equ:binary_lut1}).

The regular formula of dyadic operators of ternary logic is defined as follows:
\begin{align}
    \begin{autobreak}
        f (a_0, \cdots, a_8) 
            =        (                    x   +                     y  ) \times 1 + a_0
            \diamond (                    x   +           \overline{y} ) \times 1 + a_1
            \diamond (                    x   + \overline{\overline{y}}) \times 1 + a_2
            \diamond (          \overline{x}  +                     y  ) \times 1 + a_3
            \diamond (          \overline{x}  +           \overline{y} ) \times 1 + a_4
            \diamond (          \overline{x}  + \overline{\overline{y}}) \times 1 + a_5
            \diamond (\overline{\overline{x}} +                     y  ) \times 1 + a_6
            \diamond (\overline{\overline{x}} +           \overline{y} ) \times 1 + a_7
            \diamond (\overline{\overline{x}} + \overline{\overline{y}}) \times 1 + a_8
    \end{autobreak} \label{equ:ternary_lut2}
\end{align}
The regular formula can represent $19,683 ~ (= 3 ^ 9)$ possible dyadic operations by the combination of $a_0, \cdots, a_8$.
In the formula, one of $a_0, \cdots, a_8$ is selected by the input values $x$ and $y$, similarly, one of $a_0, \cdots, a_3$ is selected by the input values $x$ and $y$ in (\ref{equ:binary_lut2}).

\begin{table}[h]
    \centering
    \begin{tabular}{cc}
        \begin{minipage}{0.4\linewidth}
            \centering
            \caption{Truth table of ternary monadic operators} \label{tab:ternary_lut1}
            \begin{tabular}{c|c}
                    $X$ & $A$   \\
            \hline 
                    $0$ & $a_0$ \\
                    $1$ & $a_1$ \\
                    $2$ & $a_2$ \\
            \end{tabular}
        \end{minipage} &
        \begin{minipage}{0.4\linewidth}
            \centering
            \caption{Truth table of ternary dyadic operators} \label{tab:ternary_lut2}
            \begin{tabular}{cc|c}
                    $X$ & $Y$ & $A$   \\
            \hline 
                    $0$ & $0$ & $a_0$ \\
                    $0$ & $1$ & $a_1$ \\
                    $0$ & $2$ & $a_2$ \\
                    $1$ & $0$ & $a_3$ \\
                    $1$ & $1$ & $a_4$ \\
                    $1$ & $2$ & $a_5$ \\
                    $2$ & $0$ & $a_6$ \\
                    $2$ & $1$ & $a_7$ \\
                    $2$ & $2$ & $a_8$ \\
            \end{tabular}
       \end{minipage}
    \end{tabular}
\end{table}

The regular formulae also make it possible to construct operators from the truth tables shown in Table~\ref{tab:ternary_lut1} and Table~\ref{tab:ternary_lut2}.
Additionally, the regular formulae can be easily extended for operators whose number of input is $3$ or $4$, then the number of possible operators is $3 ^ {27}$ or $3 ^ {81}$.

\begin{table*}[h]
    \centering
    \caption{Truth table of dyadic operations with constant variables} \label{tab:dyadic_operations}
    \begin{tabular}{c|ccccccccc}
        $X$ & $(X \diamond 0) \diamond 0$ & $(X \diamond 0) \times 1$ & $(X \diamond 0) + 2$ & $(X \times 1) \diamond 0$ & $(X \times 1) \times 1$ & $(X \times 1) + 2$ & $(X + 2) \diamond 0$ & $(X + 2) \times 1$ & $(X + 2) + 2$ \\
    \hline 
        $0$ & $0$                         & $0$                       & $2$                  & $0$                       & $0$                     & $2$                & $2$                  & $1$                & $2$           \\
        $1$ & $0$                         & $0$                       & $2$                  & $0$                       & $1$                     & $1$                & $0$                  & $1$                & $1$           \\
        $2$ & $2$                         & $1$                       & $2$                  & $0$                       & $1$                     & $1$                & $2$                  & $1$                & $2$           \\
    \end{tabular}

    \medskip

    \begin{tabular}{c|ccccccccc}
        $X$ & $(\overline{X} \diamond 0) \diamond 0$ & $(\overline{X} \diamond 0) \times 1$ & $(\overline{X} \diamond 0) + 2$ & $(\overline{X} \times 1) \diamond 0$ & $(\overline{X} \times 1) \times 1$ & $(\overline{X} \times 1) + 2$ & $(\overline{X} + 2) \diamond 0$ & $(\overline{X} + 2) \times 1$ & $(\overline{X} + 2) + 2$ \\
    \hline 
        $0$ & $2$                                    & $1$                                  & $2$                             & $0$                                  & $1$                                & $1$                           & $2$                             & $1$                           & $2$                      \\
        $1$ & $0$                                    & $0$                                  & $2$                             & $0$                                  & $0$                                & $2$                           & $2$                             & $1$                           & $2$                      \\
        $2$ & $0$                                    & $0$                                  & $2$                             & $0$                                  & $1$                                & $1$                           & $0$                             & $1$                           & $1$                      \\
    \end{tabular}
    
    \medskip
    
    \begin{tabular}{c|ccccccccc}
        $X$ & $(\overline{\overline{X}} \diamond 0) \diamond 0$ & $(\overline{\overline{X}} \diamond 0) \times 1$ & $(\overline{\overline{X}} \diamond 0) + 2$ & $(\overline{\overline{X}} \times 1) \diamond 0$ & $(\overline{\overline{X}} \times 1) \times 1$ & $(\overline{\overline{X}} \times 1) + 2$ & $(\overline{\overline{X}} + 2) \diamond 0$ & $(\overline{\overline{X}} + 2) \times 1$ & $(\overline{\overline{X}} + 2) + 2$ \\
    \hline 
        $0$ & $0$                                               & $0$                                             & $2$                                        & $0$                                             & $1$                                           & $1$                                      & $0$                                        & $1$                                      & $1$                                 \\
        $1$ & $2$                                               & $1$                                             & $2$                                        & $0$                                             & $1$                                           & $1$                                      & $0$                                        & $1$                                      & $2$                                 \\
        $2$ & $0$                                               & $0$                                             & $2$                                        & $0$                                             & $0$                                           & $2$                                      & $2$                                        & $1$                                      & $2$                                 \\
    \end{tabular}
\end{table*}

\begin{table}[h]
    \centering
    \begin{tabular}{cc}
        \begin{minipage}{0.4\linewidth}
            \centering
            \caption{Truth table of the rotation operator} \label{tab:rot}
            \begin{tabular}{c|ccc}
                $X$ & $X$ & $\overline{X}$ & $\overline{\overline{X}}$ \\
            \hline 
                $0$ & $0$ & $2$            & $1$                       \\
                $1$ & $1$ & $0$            & $2$                       \\
                $2$ & $2$ & $1$            & $0$                       \\
            \end{tabular}
        \end{minipage} &
        \begin{minipage}{0.4\linewidth}
            \centering
            \caption{Truth table of the reverse operator} \label{tab:rev}
            \begin{tabular}{c|ccc}
                $X$ & $\widehat{X}$ & $\overline{\widehat{X}}$ & $\overline{\overline{\widehat{X}}}$ \\
            \hline 
                $0$ & $0$           & $2$                      & $1$                                 \\
                $1$ & $2$           & $1$                      & $0$                                 \\
                $2$ & $1$           & $0$                      & $2$                                 \\
            \end{tabular}
       \end{minipage}
    \end{tabular}
\end{table}

However, unlike the binary logic, some operations between a constant value and a variable can not be fully simplified.
For example, $0$ in $x \diamond 0$ can not be eliminated because $0$ is not either of the identity nor the bound of the operator $\diamond$.
However, those values contribute to extend the expressiveness of ternary calculations in exchange for their simplicity.
As shown in Table~\ref{tab:dyadic_operations}, 21 out of 27 possible monadic operations can be represented by the combinations of operators and constant values.
Table~\ref{tab:rot} and Table~\ref{tab:rev} show the rest of 6 possible operations, which can be calculated as below:
\begin{align}
    x                       &=                     x   \times 1 \diamond \overline{\overline{x}} \times 1 + 2 \label{equ:reconstruct_x}      \\
    \overline{x}            &=           \overline{x}  \times 1 \diamond                     x   \times 1 + 2 \label{equ:reconstruct_rot_x}  \\
    \overline{\overline{x}} &= \overline{\overline{x}} \times 1 \diamond           \overline{x}  \times 1 + 2 \label{equ:reconstruct_rott_x} \text{,}
\end{align}
\begin{align}
    \widehat{x}                       &=                     x   \times 1 \diamond           \overline{x}  \times 1 + 2 \label{equ:reconstruct_fx}      \\
    \overline{\widehat{x}}            &= \overline{\overline{x}} \times 1 \diamond                     x   \times 1 + 2 \label{equ:reconstruct_rot_fx}  \\
    \overline{\overline{\widehat{x}}} &=           \overline{x}  \times 1 \diamond \overline{\overline{x}} \times 1 + 2 \label{equ:reconstruct_rott_fx} \text{.}
\end{align}
Those formulae enable us to simplify the regular formula shown in (\ref{equ:ternary_lut1}) and (\ref{equ:ternary_lut2}).
The simplification process is illustrated in the next section.

\section{Results}

Conventional monadic operations such as the STI, NTI, and PTI and dyadic operations such as the TAND, TOR, TNAND, and TNOR are reconstructed with my proposed operations to illustrate the completeness of my proposal.
Moreover, a ternary half-adder circuit is designed with my proposal.

\subsection{Ternary Inverters}

\begin{table}[h]
    \centering
    \caption{Truth table of the STI, NTI and PTI} \label{tab:sti_nti_pti}
    \begin{tabular}{c|ccc}
        $X$ & $STI$ & $NTI$ & $PTI$ \\
    \hline 
        $0$ & $2$   & $2$   & $2$   \\
        $1$ & $1$   & $0$   & $2$   \\
        $2$ & $0$   & $0$   & $0$   \\
    \end{tabular}
\end{table}

\begin{figure}[h]
    \centering
    \includegraphics[width=0.5\linewidth]{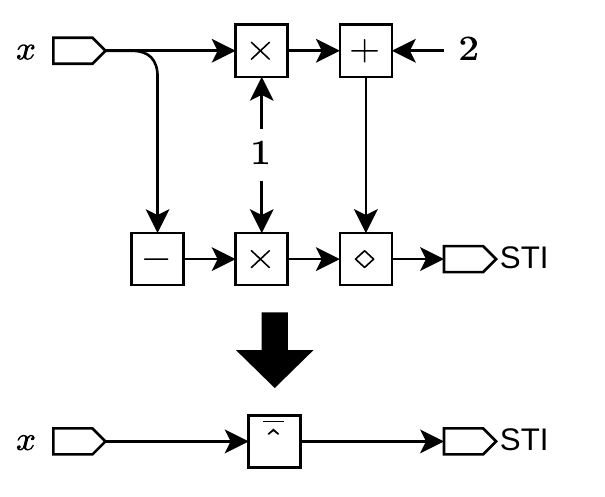}
    \caption{Standard ternary inverter (STI) circuit} \label{fig:ternary-sti}
\end{figure}

Accourding to Table~\ref{tab:sti_nti_pti} and Table~\ref{tab:ternary_lut1}, the logical expression of STI is as following:
\begin{align}
    \begin{autobreak}
        \text{STI}
            = x \times 1 + 2 \diamond \overline{x} \times 1 + 1 \diamond \overline{\overline{x}} \times 1 + 0
            = x \times 1 + 2 \diamond 1 \diamond \overline{\overline{x}} \times 1
            = \overline{\overline{x}} \times 1 \diamond x \times 1 + 2
            = \overline{\widehat{x}}
    \end{autobreak} \text{.} \label{equ:circuit_sti}
\end{align}
In this equation, $\overline{x} \times 1 + 1$ is replaced to $1$ because $1$ is the bound of $+$, and entire the term is emitted because $1$ also works as the identity element of $\diamond$.
As the result, the equations becomes the same as (\ref{equ:reconstruct_rot_fx}).
Figure~\ref{fig:ternary-sti} shows the generated circuit.
As shown in the figure, the circuit is composed of five operators.

\begin{figure}[h]
    \centering
    \includegraphics[width=0.5\linewidth]{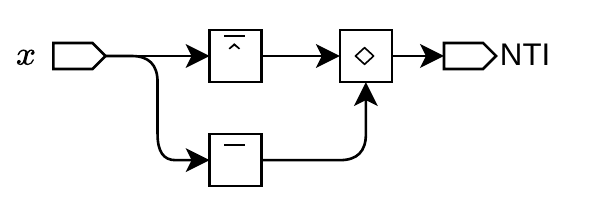}
    \caption{Negative ternary inverter (NTI) circuit} \label{fig:ternary-nti}
\end{figure}

The expression that corresponds to NTI is shown below:
\begin{align}
    \begin{autobreak}
        \text{NTI}
            = x \times 1 + 2 \diamond \overline{x} \times 1 + 0 \diamond \overline{\overline{x}} \times 1 + 0
            = x \times 1 + 2 \diamond x \times 1 + 2 \diamond \overline{x} \times 1 \diamond \overline{\overline{x}} \times 1
            = (\overline{\overline{x}} \times 1 \diamond x \times 1 + 2) \diamond (\overline{x} \times 1 \diamond x \times 1 + 2)
            = \overline{\widehat{x}} \diamond \overline{x}
    \end{autobreak} \text{.} \label{equ:circuit_nti}
\end{align}
The idempotency law allows us to duplicate a term $x \times 1 + 2$, then two terms $\overline{\overline{x}} \times 1 \diamond x \times 1 + 2$ and $\overline{x} \times 1 \diamond x \times 1 + 2$ are made due to the associativity law.
The replacement is done by (\ref{equ:reconstruct_rot_fx}) and (\ref{equ:reconstruct_rott_x}).
Figure~\ref{fig:ternary-nti} illustrates the generated circuit.
As shown in the figure, the circuit includes three operators which include standard ternary inverter.

\begin{figure}[h]
    \centering
    \includegraphics[width=0.5\linewidth]{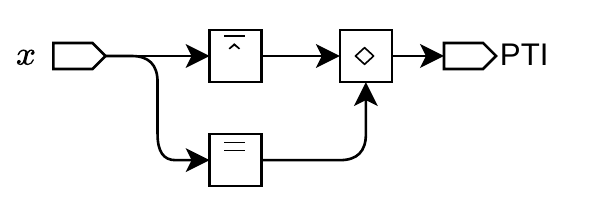}
    \caption{Positive ternary inverter (PTI) circuit} \label{fig:ternary-pti}
\end{figure}

The expression that corresponds to PTI is shown below:
\begin{align}
    \begin{autobreak}
        \text{PTI}
            = x \times 1 + 2 \diamond \overline{x} \times 1 + 2 \diamond \overline{\overline{x}} \times 1 + 0
            = x \times 1 + 2 \diamond \overline{x} \times 1 + 2 \diamond \overline{\overline{x}} \times 1 \diamond \overline{\overline{x}} \times 1
            = (\overline{\overline{x}} \times 1 \diamond x \times 1 + 2) \diamond (\overline{\overline{x}} \times 1 \diamond \overline{x} \times 1 + 2)
            = \overline{\widehat{x}} \diamond \overline{\overline{x}}
        \end{autobreak} \text{.} \label{equ:circuit_pti}
\end{align}
The calculation procedure is similar to the case of NTI.
The positive ternary inverter circuit is shown in Fig.~\ref{fig:ternary-pti}.

\subsection{Ternary NAND (TNAND) and Ternary NOR (TNOR)}

\begin{table}[h]
    \centering
    \caption{Truth table of the TNAND and TNOR} \label{tab:tnand_tnor}
    \begin{tabular}{cc|cc}
        $X$ & $Y$ & $TNAND$ & $TNOR$ \\
    \hline 
        $0$ & $0$ & $2$     & $2$    \\
        $0$ & $1$ & $2$     & $1$    \\
        $0$ & $2$ & $2$     & $0$    \\
        $1$ & $0$ & $2$     & $1$    \\
        $1$ & $1$ & $1$     & $1$    \\
        $1$ & $2$ & $1$     & $0$    \\
        $2$ & $0$ & $2$     & $0$    \\
        $2$ & $1$ & $1$     & $0$    \\
        $2$ & $2$ & $0$     & $0$    \\
    \end{tabular}
\end{table}

\begin{figure}
    \centering
    \includegraphics[width=0.7\linewidth]{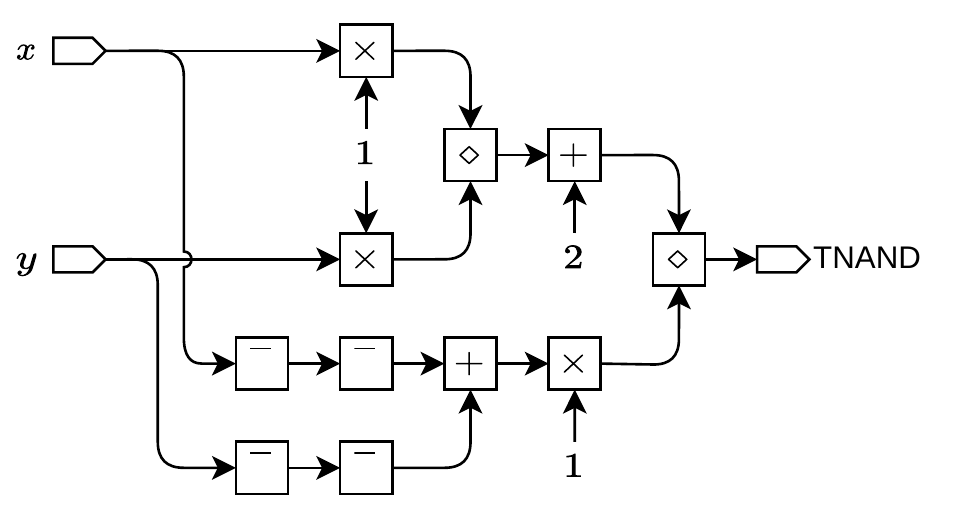}
    \caption{Ternary NAND (TNAND) circuit} \label{fig:ternary-nand}
\end{figure}

The logical expression of the TNAND generated by Table~\ref{tab:ternary_lut2} is as follows:
\begin{align}
    \begin{autobreak}
        \text{TNAND}
            = (x + y) \times 1 + 2 \diamond (x + \overline{y}) \times 1 + 2 \diamond (x + \overline{\overline{y}}) \times 1 + 2
            \diamond (x + y) \times 1 + 2 \diamond (\overline{x} + y) \times 1 + 2 \diamond (\overline{\overline{x}} + y) \times 1 + 2
            \diamond (\overline{\overline{x}} + \overline{\overline{y}}) \times 1
            = (x \times 1 + y \times 1 \diamond x \times 1 + \overline{y} \times 1 \diamond x \times 1 + \overline{\overline{y}} \times 1) + 2
            \diamond (x \times 1 + y \times 1 \diamond \overline{x} \times 1 + y \times 1 \diamond \overline{\overline{x}} \times 1 + y \times 1) + 2
            \diamond (\overline{\overline{x}} + \overline{\overline{y}}) \times 1
            = (y \times 1 \diamond \overline{y} \times 1 \diamond \overline{\overline{y}} \times 1) + x \times 1 + 2
            \diamond (x \times 1 \diamond \overline{x} \times 1 \diamond \overline{\overline{x}} \times 1) + y \times 1 + 2
            \diamond (\overline{\overline{x}} + \overline{\overline{y}}) \times 1
            = x \times 1 + 2 \diamond y \times 1 + 2 \diamond (\overline{\overline{x}} + \overline{\overline{y}}) \times 1
            = (x \times 1 \diamond y \times 1) + 2 \diamond (\overline{\overline{x}} + \overline{\overline{y}}) \times 1
    \end{autobreak} \text{.} \label{equ:circuit_tnand}
\end{align}
In the equation, the terms which include $+ 1$ are already omitted and term duplication is done.
As the result, all terms are roughly classified into three parts the first one involves $x$, the second one involves $y$, and the other one.
The terms are expanded by the distributivity law, and then the first and the second classified terms can be bundled by $+ 2$ then, the first and second terms inside $+ 2$ can be bundled by $x \times 1$, $y \times 1$ respectively.
The part $y \times 1 \diamond \overline{y} \times 1 \diamond \overline{\overline{y}} \times 1$ and $x \times 1 \diamond \overline{x} \times 1 \diamond \overline{\overline{x}} \times 1$ are calculated to be $0$ as shown below:
\begin{align}
    \begin{autobreak}
        x \times 1 \diamond \overline{x} \times 1 \diamond \overline{\overline{x}} \times 1 
            = (x \diamond \overline{x} \diamond \overline{\overline{x}})
            \times (x \diamond \overline{x}) \times (\overline{x} \diamond \overline{\overline{x}}) \times (\overline{\overline{x}} \diamond x)
            \times (x \times \overline{x} \times \overline{\overline{x}}) \times 1
            = 2
            \times (x \diamond \overline{x}) \times (\overline{x} \diamond \overline{\overline{x}}) \times (\overline{\overline{x}} \diamond x)
            \times 0 \times 1
            = 0
    \end{autobreak} \text{.}
\end{align}
The generated TNAND circuit is shown in Fig.~\ref{fig:ternary-nand}.

\begin{figure}
    \centering
    \includegraphics[width=0.7\linewidth]{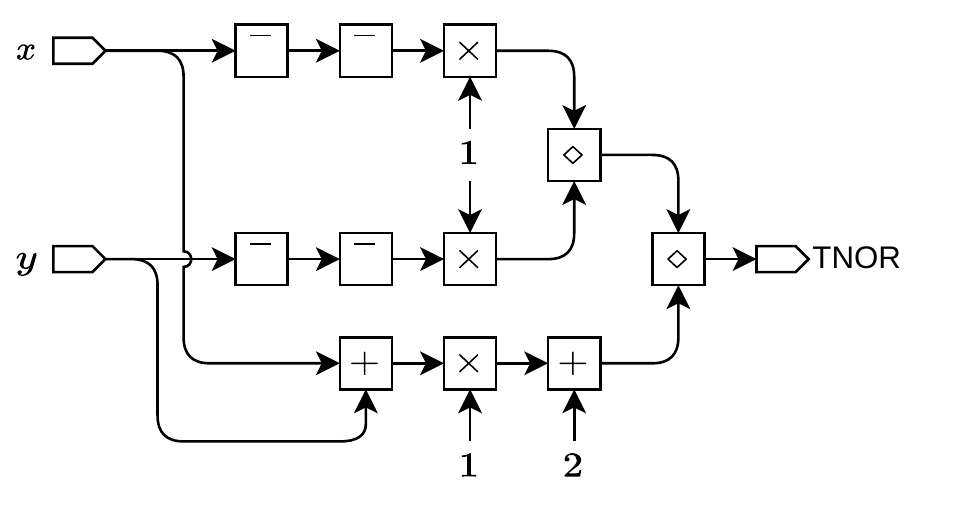}
    \caption{Ternary NOR (TNOR) circuit} \label{fig:ternary-nor}
\end{figure}

The logical expression of TNOR generated by Table~\ref{tab:ternary_lut2} is shown in the following equation:
\begin{align}
    \begin{autobreak}
        \text{TNOR}
            = (\overline{\overline{x}} + y) \times 1 \diamond (\overline{\overline{x}} + \overline{y}) \times 1 \diamond (\overline{\overline{x}} + \overline{\overline{y}}) \times 1
            \diamond (x + \overline{\overline{y}}) \times 1 \diamond (\overline{x} + \overline{\overline{y}}) \times 1 \diamond (\overline{\overline{x}} + \overline{\overline{y}}) \times 1
            \diamond (x + y) \times 1 + 2
            = \overline{\overline{x}} \times 1 + y \times 1 \diamond \overline{\overline{x}} \times 1 + \overline{y} \times 1 \diamond \overline{\overline{x}} \times 1 + \overline{\overline{y}} \times 1
            \diamond x \times 1 + \overline{\overline{y}} \times 1 \diamond \overline{x} \times 1 + \overline{\overline{y}} \times 1 \diamond \overline{\overline{x}} \times 1 + \overline{\overline{y}} \times 1
            \diamond (x + y) \times 1 + 2
            = (y \times 1 \diamond \overline{y} \times 1 \diamond \overline{\overline{y}} \times 1) + \overline{\overline{x}} \times 1
            \diamond (x \times 1 \diamond \overline{x} \times 1 \diamond \overline{\overline{x}} \times 1) + \overline{\overline{y}} \times 1
            \diamond (x + y) \times 1 + 2
            = \overline{\overline{x}} \times 1 \diamond \overline{\overline{y}} \times 1 \diamond (x + y) \times 1 + 2
    \end{autobreak} \text{.} \label{equ:circuit_tnor}
\end{align}
The equation results in simple operations by the calculation as TNAND, and its circuit is shown in Fig.~\ref{fig:ternary-nor}.
The logical expressions of TAND and TOR operation can also be obtained by (\ref{equ:ternary_lut2}) as this paper demonstrated with TNAND and TOR.

\subsection{Ternary Half-Adder (THA)}

\begin{table}[h]
    \centering
    \caption{Truth table of a ternary half-adder} \label{tab:ternary_half_adder}
    \begin{tabular}{cc|cc}
        $X$ & $Y$ & $Carry$ & $Sum$ \\
    \hline 
        $0$ & $0$ & $0$     & $0$   \\
        $0$ & $1$ & $0$     & $1$   \\
        $0$ & $2$ & $0$     & $2$   \\
        $1$ & $0$ & $0$     & $1$   \\
        $1$ & $1$ & $0$     & $2$   \\
        $1$ & $2$ & $1$     & $0$   \\
        $2$ & $0$ & $0$     & $2$   \\
        $2$ & $1$ & $1$     & $0$   \\
        $2$ & $2$ & $1$     & $1$   \\
    \end{tabular}
\end{table}

\begin{figure}
    \centering
    \includegraphics[width=0.8\linewidth]{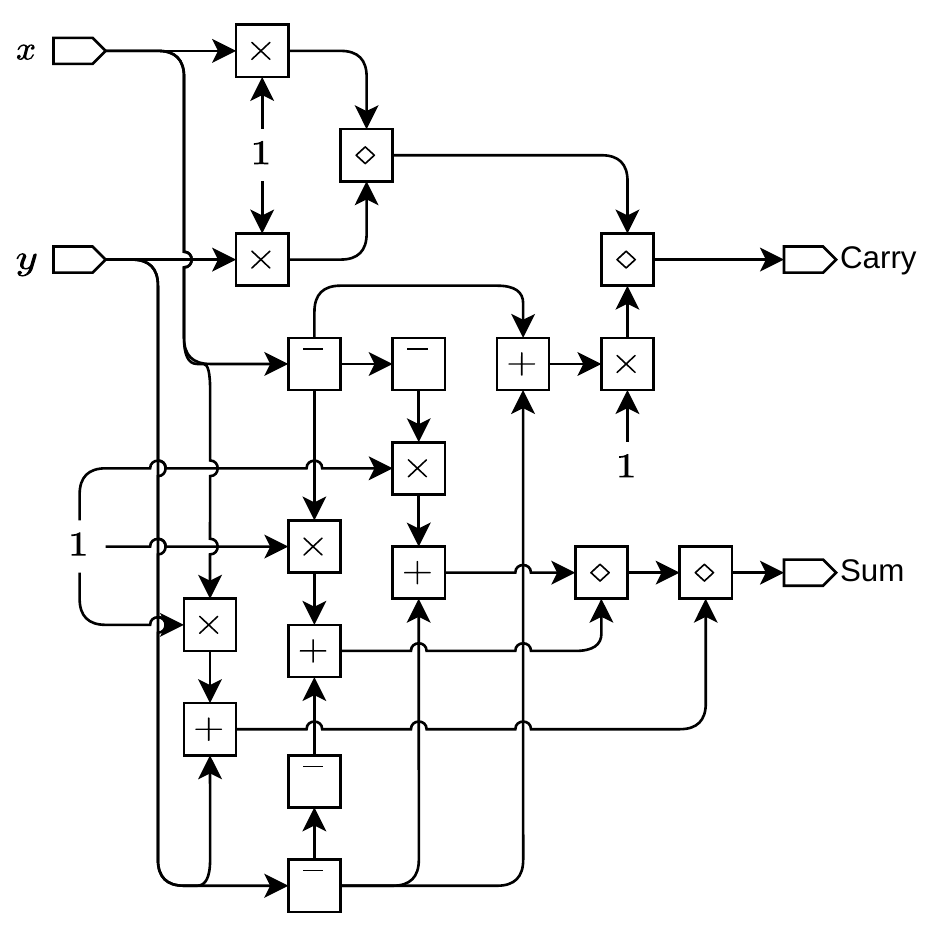}
    \caption{Ternary half-adder (THA) circuit} \label{fig:ternary-half-adder}
\end{figure}

The THA also can be constructed in my proposal.
The THA is composed of two signals: Sum and Carry.
The logical expression of $\text{Carry}$ and $\text{Sum}$ are as following:
\begin{align}
    \begin{autobreak}
        \text{Carry}
            = (x + y) \times 1 \diamond (x + \overline{y}) \times 1 \diamond (x + \overline{\overline{y}}) \times 1
            \diamond (\overline{x} + y) \times 1 \diamond (\overline{x} + \overline{y}) \times 1 \diamond (\overline{\overline{x}} + y) \times 1
            = x \times 1 + y \times 1 \diamond x \times 1 + \overline{y} \times 1 \diamond x \times 1 + \overline{\overline{y}} \times 1
            \diamond \overline{x} \times 1 + y \times 1 \diamond \overline{x} \times 1 + \overline{y} \times 1 \diamond \overline{\overline{x}} \times 1 + y \times 1
            = (y \times 1 \diamond \overline{y} \times 1 \diamond \overline{\overline{y}} \times 1) + x \times 1
            \diamond (x \times 1 \diamond \overline{x} \times 1 \diamond \overline{\overline{x}} \times 1) + y \times 1
            \diamond (\overline{x} + \overline{y}) \times 1
            = x \times 1 \diamond y \times 1 \diamond (\overline{x} + \overline{y}) \times 1
    \end{autobreak} \text{,} \label{equ:circuit_carry}
\end{align}
\begin{align}
    \begin{autobreak}
        \text{Sum}
            = (x + y) \times 1 \diamond (x + \overline{\overline{y}}) \times 1 + 2
            \diamond (\overline{x} + \overline{\overline{y}}) \times 1 \diamond (\overline{x} + \overline{y}) \times 1 + 2
            \diamond (\overline{\overline{x}} + y) \times 1 + 2 \diamond (\overline{\overline{x}} + \overline{y}) \times 1
            = x \times 1 + y \times 1 \diamond x \times 1 +\overline{\overline{y}} \times 1 + 2
            \diamond \overline{x} \times 1 + \overline{\overline{y}} \times 1 \diamond \overline{x} \times 1 + \overline{y} \times 1 + 2
            \diamond \overline{\overline{x}} \times 1 + \overline{y} \times 1 \diamond \overline{\overline{x}} \times 1 + y \times 1 + 2
            = x \times 1 + (y \times 1 \diamond \overline{\overline{y}} \times 1 + 2)
            \diamond \overline{x} \times 1 + (\overline{\overline{y}} \times 1 \diamond \overline{y} \times 1 + 2)
            \diamond \overline{\overline{x}} \times 1 + (\overline{y} \times 1 \diamond y \times 1 + 2)
            = x \times 1 + y \diamond \overline{x} \times 1 + \overline{\overline{y}} \diamond \overline{\overline{x}} \times 1 + \overline{y}
    \end{autobreak} \text{.} \label{equ:circuit_sum}
\end{align}
The part $y \times 1 \diamond \overline{y} \times 1 \diamond \overline{\overline{y}} \times 1$ and $x \times 1 \diamond \overline{x} \times 1 \diamond \overline{\overline{x}} \times 1$ are simplified to $0$ as shown in a calculation step of TNAND.
Moreover, the part $y \times 1 \diamond \overline{\overline{y}} \times 1 + 2$, $\overline{\overline{y}} \times 1 \diamond \overline{y} \times 1 + 2$, and $\overline{y} \times 1 \diamond y \times 1 + 2$ can be replaced with $y$, $\overline{\overline{y}}$, and $\overline{y}$ by (\ref{equ:reconstruct_x}), (\ref{equ:reconstruct_rot_x}), and (\ref{equ:reconstruct_rott_x}), respectively.
The ternary circuit of the THA is illustrated in Fig.~\ref{fig:ternary-half-adder}.

\section{Discussion}

This paper proposed new symmetric ternary operators and systematic design methodology of ternary logic circuits.
The results in this work demonstrated that any ternary calculation circuits can be designed by the four essential ternary operators named ROTATE, ALPHA, BETA, and GAMMA.
The combination of those ternary operators achieved a full-ternary implementation of the THA circuit meanwhile conventional implementations includes binary logic gates~\cite{lin2009cntfet,jaber2019high}.

However, the increase in the number of inputs of the ternary truth table results in a huge calculation cost.
The simplification of the general formula requires complicated calculation due to its three different dyadic operators.
A systematic simplification algorithm is required to exploit the potential of my proposal.
For instance, NTI and PTI can also be represented as $\overline{X} \diamond 0$ and $(\overline{X} + 2) \diamond 0$ although this work did not figure out the transformations.

Additionally, a systematic transformation algorithm from the traditional binary logic into my proposed ternary logic is desired.
This transformation and simplification algorithms realize performance enhancement and area reduction of the current VLSI.

\bibliographystyle{IEEEtran}
\nocite{*}
\bibliography{Main.bbl}

\end{document}